\documentclass[10pt,twocolumn,english,nofootinbib,twocolumn]{revtex4-1}
\usepackage{ae,aecompl}
\usepackage[T1]{fontenc}
\usepackage[latin9]{inputenc}
\usepackage{bm}
\usepackage{amsmath}
\usepackage{amssymb}
\usepackage{graphicx}
\usepackage{esint}
\usepackage{array}
\usepackage{booktabs }
\usepackage{topcapt}

\usepackage[unicode,breaklinks]{hyperref}
\hypersetup{
    unicode=true,
    a4paper=true,
    plainpages=false, 
    colorlinks=true,
    linkcolor=blue,
    citecolor=blue,
    filecolor=black,
    urlcolor=blue
}
\newcolumntype{C}{>{$\displaystyle} c <{$}}

\makeatletter
\usepackage{bbold}
\usepackage{calrsfs}

\makeatother
\newcommand{\bk}{\mathbf{k}}
\newcommand{\bu}{\mathbf{u}}
\newcommand{\bv}{\mathbf{v}}
\newcommand{\kt}{k_B T}

\newcommand{\qt}{\tilde{q}}

\newcommand{\fh}{\hat{f}}
\newcommand{\nh}{\hat{n}}
\newcommand{\wh}{\hat{w}}
\newcommand{\fracb}[2]{\left(\frac{#1}{#2}\right)}
\newcommand{\sqrtf}[2]{\sqrt{\frac{#1}{#2}}}
\newcommand{\efree}{\epsilon_\bk}

\usepackage{babel}
\begin{document}

\title{Static structure factors for a spin-1 Bose-Einstein condensate}

\author{L. M. Symes}

\author{D. Baillie}

\author{P. B. Blakie}

\affiliation{Jack Dodd Centre for Quantum Technology, Department of Physics, University
of Otago, Dunedin, New Zealand}
\begin{abstract}
We consider the total density and spin density fluctuations of a uniform spin-1
Bose-Einstein condensate within the Bogoliubov formalism. We present
 results for the total density and spin density static structure
factors for all four magnetic phases. A key result of our work is a set of analytic predictions for the structure factors in the 
large and small momentum limits.
These results will be useful in current experiments aiming to develop a better understanding of the excitations and fluctuations of spinor condensates. 
\end{abstract}
\maketitle

\section{Introduction}
A spinor Bose-Einstein condensate (BEC) consists of atoms with a spin degree of freedom  \cite{Ho1998a,Ohmi1998a}. In addition to exhibiting spatial coherence, a spinor condensate also displays a range of spin orders, determined by the interactions and externally applied magnetic field. Various aspects of the phase diagram and condensate dynamics have been explored in experiments, particularly for the case of spin-1 where the atoms can access three magnetic sublevels (e.g.~see \cite{Stenger1999a,Chang2004a,Chang2005a,Black2007a,Vengalattore2008a,Vengalattore2010a,Liu2009b,Sadler2006a,Liu2009a}). An important feature of this system is that it exhibits a rich excitation spectrum with phonon and magnon branches \cite{Ho1998a,Ohmi1998a,Kawaguchi2012R,Stamper-Kurn2013R}. 

In this paper we develop a formalism to describe the fluctuations of the various densities of interest for a spin-1 condensate. Our primary focus is 
the total number density and the components of the spin density, motivated by the capability to measure these quantities directly in experiments (e.g.~by Stern-Gerlach \cite{Bookjans2011a,Lucke2011a,Hamley2012a,Vinit2013a} and dispersive  \cite{Carusotto2004a, Liu2009a, Higbie2005a,Vengalattore2010a,Eckert2007a} probing).
We characterise these fluctuations by calculating the relevant 
static structure factors.
The Bogoliubov description of the spin-1 condensate is expected to provide a good description of the system for temperatures well below the condensation temperature. Within this framework, we present both numerical results and analytic expressions for the limiting behavior of the static structure factors.  
For each of the four distinct magnetic phases of the spin-1 condensate, we relate how the three Bogoliubov excitation branches contribute to the  fluctuations.  
Of particular interest are the antiferromagnetic and broken-axisymmetric phases, in which a second continuous symmetry associated with the spin degree of freedom is broken [in addition to the $U(1)$ gauge symmetry]. This is revealed by the emergence of a second Nambu-Goldstone mode \cite{Uchino2010a}.

For the case of the total density,  
the long wavelength limit of the structure factor is 
$k_BT/Mc_n^2$ 
where $T$ is the temperature, $M$ is the atomic mass and $c_n$ is the speed of sound (also see \cite{Uchino2010a}).
This is equivalent to the thermodynamic result $\Delta N^2=Vn^2 k_BT\kappa$, where $\Delta N^2$ is the number variance in a volume $V$ of a system of average density $n$ with  isothermal compressibility $\kappa=1/nMc_n^2$.  
We also  analyse the structure factors for the three components of spin density.  
Analogous to the relation between fluctuations and compressibility for 
the density static structure factor,  
the long wavelength limit of the spin density structure factors reveals the magnetic susceptibility of the condensate.

While the dynamic and static structure factors are well characterised for the case of scalar condensates (e.g.~see \cite{Zambelli2000}), much less work has been done on multicomponent systems, although we note theoretical studies of binary condensates \cite{Chung2008a,Abad2013a} and an approximate treatment of the finite temperature transverse spin-density correlations in a quasi-two-dimensional ferromagnetic condensate (see Appendix B of Ref.~\cite{Barnett2011a}). Experimentally the static structure factor can be determined directly from fluctuation measurements  (e.g.~see \cite{Hung2011,Blumkin2013a}), off-resonant light scattering \cite{Sykes2011a} and Bragg spectroscopy \cite{Steinhauer2002a,Kuhnle2010a}. Notably, in recent experiments spin-dependent Bragg spectroscopy has been used to measure the $z$-spin density of a  spin-$\frac{1}{2}$ Fermi gas \cite{Hoinka2012a}, and  speckle imaging has  been employed to measure the compressibility and magnetic susceptibility of a strongly interacting Fermi gas \cite{Sanner2011a}. Along this path a number of  experiments with spin-1 condensates have made fluctuation measurements, particularly in application to dynamical regimes (e.g.~\cite{Liu2009b,Guzman2011a,Bookjans2011a}) and spin-squeezing \cite{Hamley2012a}. We also note a recent proposal to use magnetic spectroscopy to impart energy to a spinor condensate for the purposes of probing its excitation spectrum \cite{Tokuno2013a}.

The structure of this paper is as follows. In Sec.~\ref{Sec:System} we introduce the Hamiltonian and meanfield description of the spin-1 system. We present the phase diagram and briefly discuss the four distinct equilibrium phases. In Sec.~\ref{Sec:flucts} we present a general treatment of fluctuations in the spin-1 system by introducing a generalised two-point density correlation function, from which we obtain the static structure factors. In Sec.~\ref{Sec:Res1}  we discuss the excitation spectrum and the relationship of each branch of the spectrum  to the fluctuations of interest for each of the four equilibrium phases. We present both numerical and analytic results for the various static structure factors. The analytic results are summarised in Table \ref{tab:Slimits}. Finally, we conclude our work in Sec.~\ref{Sec:Discussion-Conclusion}, discussing the possible applications  of our results.
 
\section{System}\label{Sec:System}

\subsection{Hamiltonian}
We consider a uniform three-dimensional spin-1 Bose gas  subject to a uniform magnetic
field along $z$.  
The single-particle description of the atoms is provided by the Hamiltonian
\begin{equation}
(h_{0})_{ij}=\left[-\frac{\hbar^{2}\nabla^{2}}{2M}-pi+qi^{2}\right]\delta_{ij},
\end{equation}
 where $p$ and $q$ are the coefficients of the linear\footnote{The quantity $p$  also serves as  a Lagrange multiplier  to constrain the $z$ component of magnetization.} and quadratic
Zeeman terms, respectively,  and the subscripts $i,j=\{-,0,+\}$ refer
to the $m_{F}=\{-1,0,1\}$ magnetic sub-levels of the atoms. The value of $q$ is tunable independently of
$p$ (e.g.~see \cite{Gerbier2006,Bookjans2011b}) and can be both positive and negative.

The cold-atom Hamiltonian, including interactions, is given by \cite{Ho1998a,Ohmi1998a}
\begin{align}
\hat{H}\!=\! & \int\! d \mathbf{x}\, \hat{\boldsymbol{\psi}}^{\dagger}(\mathbf{x})h_{0}\hat{\bm{\psi}}(\mathbf{x})+:\!\frac{c_{0}}{2}\hat{n}(\mathbf{x})\hat{n}(\mathbf{x})+\frac{c_{1}}{2}\hat{\mathbf{f}}(\mathbf{x})\cdot\hat{\mathbf{f}}(\mathbf{x})\!:, \label{eq:H_org}
\end{align}
 where  $::$ indicates normal ordering, 
$\hat{\bm{\psi}}=[\hat{\psi}_{+},\hat{\psi}_{0},\hat{\psi}_{-}]^{T}$
is the spinor boson field operator, and the superscript $T$ indicates the transpose operation. The interaction terms involve
the total density $\hat{n}$ and the spin density $\hat{\mathbf{f}}=[\hat{f}_x,\hat{f}_y,\hat{f}_z]^T$
given by 
\begin{eqnarray}
\hat{n}(\mathbf{x}) & = & \hat{\boldsymbol{\psi}}^{\dagger}\!(\mathbf{x)}\hat{\boldsymbol{\psi}}(\mathbf{x)},\label{eq:ndef}\\
\hat{f}_{\alpha}(\mathbf{x}) & = & \hat{\boldsymbol{\psi}}^{\dagger}\!(\mathbf{x})\mathrm{F}_{\alpha}\hat{\boldsymbol{\psi}}(\mathbf{x}),\quad\alpha=x,y,z,\label{eq:Fdef}
\end{eqnarray}
where $\{\mathrm{F}_{\alpha}\}$
are the spin-1 matrices. The parameters $c_0$ and $c_1$ are  the density and spin dependent interaction parameters,
respectively, and are given by $c_0=4\pi\hbar^{2}(a_{0}+2a_{2})/3M$ and
$c_1=4\pi\hbar^{2}(a_{2}-a_{0})/3M$, with $a_{S}$ ($S=0,2$) being
the $s$-wave scattering length for the scattering channel of total
spin $S$.

\subsection{Meanfield description of system}
Here we shall be interested in temperatures well below the condensation temperature, where the field can be written as 
\begin{equation}
\hat{\boldsymbol{\psi}}(\mathbf{x})=\sqrt{n}\,\boldsymbol{\xi}+\hat{\boldsymbol{\delta}}(\mathbf{x}),\label{eq:psi-1}
\end{equation}
where $\langle\hat{\boldsymbol{\psi}}\rangle=\sqrt{n}\,\boldsymbol{\xi}$
is the (uniform) condensate field,   $n=N/V$ is the condensate
density, $V$ is the volume, $N$ is the number of condensate atoms, and $\boldsymbol{\xi}=[\xi_{+},\xi_{0},\xi_{-}]^{T}$ is the normalized
condensate spinor. The operator $\hat{\boldsymbol{\delta}}=[\hat{\delta}_{+},\hat{\delta}_{0},\hat{\delta}_{-}]^{T}$
represents the non-condensate field. 

\subsubsection{Condensate and phase diagram}
\begin{figure}
\includegraphics[width=3.5in]{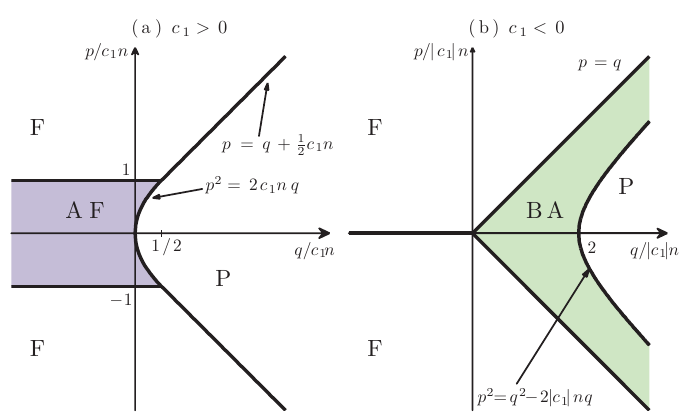}
\caption{\label{fig:PD}The zero temperature phase diagram of a spin-1 Bose
gas for cases with  (a) \emph{antiferromagnetic} interactions 
(i.e.~$c_1>0$), and (b) \emph{ferromagnetic} interactions 
(i.e.~$c_1<0$). The vertical
and horizontal axes are the linear and quadratic Zeeman energies (see
text) in units of  $|c_{1}|n$, where $n$ is the condensate number
density. The phases shown are (F) ferromagnetic, (P) polar, (AF) antiferromagnetic,
 and broken-axisymmetric (BA) (see Refs. \cite{Stenger1999a,Kawaguchi2012R}).
The rotational symmetry about the direction of the applied field is
spontaneously broken in the AF and BA phases.}
\end{figure}

The condensate is obtained as the lowest energy solution of the Gross-Pitaevskii
equation
\begin{equation}
\mu\boldsymbol{\xi}=\left[h_{0}+c_0 n\mathbb{1}+c_1\sum_{\alpha}f_\alpha\mathrm{F}_{\alpha}\right]\boldsymbol{\xi},\label{GPE}
\end{equation}
where $\mathbb{1}$ is the $3\times3$ identity matrix and 
\begin{equation}
f_\alpha= n\boldsymbol{\xi}^{\dagger}\mathrm{F}_{\alpha}\boldsymbol{\xi},
\end{equation}
is the $\alpha$-component of the condensate spin density. 
A variety of ground state phases emerge from the competition between the spin-dependent interaction (i.e. $c_1n$) and the external magnetic field (i.e.~$p$ and $q$). For spin-1 there are four distinct phases distinguished by their magnetization, both along the direction of the external field (i.e.~$f_z$) and perpendicular to it (i.e.~$f_\perp \equiv\sqrt{f_x^2+f_y^2}$). These properties are summarized in Table \ref{tab:phases}, and the parameter  regions where each phase is the predicted ground state is shown in  Fig.~\ref{fig:PD}.

\begin{table}[htbp]
   \centering  
  {\small  \begin{tabular}{ p{3.8cm} p{4.6cm}  } 
    \hline
     Phase   & Properties \\ \hline
    {\bf Ferromagnetic (F)} &  Fully magnetized   $|f_z|=n$, $|f_\perp|=0$.  $\boldsymbol{\xi}=[1,0,0]^T$ or $[0,0,1]^T$. \\[0.15cm]
    {\bf Polar (P)} &  Unmagnetized  $|f_z|=|f_\perp|=0$. $\boldsymbol{\xi}=[0,1,0]^T$   \\[0.15cm]  
  {\bf Anti-ferromagnetic (AF)} &  Partially magnetized  $|f_z|\le n$, $|f_\perp|=0$. Condensate spinor has non-zero components in the $m_F=\pm1$ sublevels. \\[0.15cm]
    {\bf Broken-axisymmetric (BA)} & Partially magnetized, but tilts to the $z$ axis giving $f_\perp>0$. Condensate spinor has non-zero components in all  sublevels. \\[0.1cm]
    
 \hline\hline
    \end{tabular}
 \caption{The  phases of a spin-1 BEC, as  presented in Fig.~\ref{fig:PD}, categorised according to their magnetization.  \label{tab:phases}
 } }
\end{table}

Detailed derivations of the ground states and the phase diagram are too lengthy to present here, and we refer the reader to the excellent summary given in Sec.~3.3 of Ref.~\cite{Kawaguchi2012R}.
We also note here that, in addition to the spin density, an important characterization of the condensate order is provided by the nematic tensor  
\begin{equation}
 {q}_{\alpha\beta} = n\boldsymbol{\xi}^{\dagger} \mathrm{Q}_{\alpha\beta}  \boldsymbol{\xi},\label{eq:nematic_cond}
\end{equation}
where $\mathrm{Q}_{\alpha\beta} =\tfrac{1}{2}\left({\mathrm{F}_\alpha \mathrm{F}_\beta +\mathrm{F}_\beta \mathrm{F}_\alpha}  \right)$ is a 3$\times$3 matrix for each pair of $\alpha,\beta\in\{x,y,z\}$.

\subsubsection{Bogoliubov excitations}
The excitations of the condensate are determined by the non-condensate
operator. Within a Bogoliubov treatment this operator can be expressed
as 
\begin{equation}
\hat{\boldsymbol{\delta}}(\mathbf{x})=\sum_{\mathbf{k}\ne\mathbf0,\nu}(\mathbf{u}_{\mathbf{k}\nu}\hat{\alpha}_{\mathbf{k}\nu}-\mathbf{v}_{\mathbf{k}\nu}^{*}\hat{\alpha}_{-\mathbf{k}\nu}^{\dagger})\frac{e^{i\mathbf{k}\cdot\mathbf{x}}}{\sqrt{V}},
\end{equation}
 where $\{\bu_{\bk\nu},\bv_{\bk\nu}\}$
are the quasiparticle amplitudes, with respective energies $E_{\mathbf{k}\nu}$,
and $\nu=\{0,1,2\}$ is the spin mode label distinguishing the three solution branches. The quasiparticle operators $\hat{\alpha}_{\mathbf{k}\nu}$ satisfy bosonic commutation relations. 

Quite a broad understanding of the quasiparticle solutions has been developed for the spin-1 condensate, however the full review of this is too lengthy to be included here, and we refer the reader to Refs.~\cite{Murata2007a,Uchino2010a,Kawaguchi2012R}. 
We make use of a number of these results in the expressions we derive here for the static structure factors. The results which we present are obtained by diagonalising a $6\times6$ matrix to determine the  quasiparticle energies and amplitudes for the three branches (e.g.~see Secs.~5.1 and 5.2 of Ref.~\cite{Kawaguchi2012R}). This is done for each $k$ for the numerical results and analytically for the results in Table~\ref{tab:Slimits}.

\section{Fluctuations}\label{Sec:flucts}

\subsection{Observable}
Our interest lies in the fluctuations that occur in the total and spin
densities of the system, as characterized by the observables given
in Eqs. (\ref{eq:ndef}) and (\ref{eq:Fdef}). We generically
represent these observables as
\begin{equation}
\hat{w}(\mathbf{x})=\hat{\boldsymbol{\psi}}^{\dagger}\!(\mathbf{x)}\mathrm{W}\hat{\boldsymbol{\psi}}(\mathbf{x)},\label{eq:Qdef}
\end{equation}
where   $\mathrm{W}$ is a 3$\times$3 matrix.\footnote{In this paper we consider the cases of $\mathrm{W}\in\{\mathbb{1},\mathrm{F}_x,\mathrm{F}_y,\mathrm{F}_z\}$, i.e.~$\hat{w}(\mathbf{x})$ being the total density or a component of the spin density.} 
In the low-temperature regime of interest the mean value is determined by the condensate and is spatially
constant, i.e.
\begin{equation}
 {w}=\langle\hat{w}(\mathbf{x})\rangle=n\boldsymbol{\xi}^{\dagger}\mathrm{W}\boldsymbol{\xi},
\end{equation}
and in what follows we consider the fluctuations about this mean value.

\subsection{$w$ density-density correlation function}
The spatial fluctuations of $\hat{w}$   are characterized by the two-point correlation function 
\begin{equation}
C_{w}(\mathbf{x}-\mathbf{x}')=\left\langle\delta \hat w(\mathbf{x})\delta \hat w(\mathbf{x}')\right\rangle,
\end{equation}
where we have introduced the  fluctuation operator
\begin{equation}
\delta \hat w(\mathbf{x})=\hat{w}(\mathbf{x})- {w}.
\end{equation}
Because we consider a uniform system, $C_w$ only depends on the relative separation of the two points.

It is convenient to rewrite the correlation function in the form 
\begin{equation}
C_{w}(\mathbf{x}-\mathbf{x}')=\left\langle:\delta \hat w(\mathbf{x})\delta \hat w(\mathbf{x}'):\right\rangle +\overline{w^2}\,\delta(\mathbf{x}-\mathbf{x}'),\label{CQnormal}
\end{equation}
where
\begin{equation}
\overline{w^2}\equiv \langle \hat{\boldsymbol{\psi}}^{\dagger}\!(\mathbf{x)}\mathrm{W}^2\hat{\boldsymbol{\psi}}(\mathbf{x)}\rangle=n\boldsymbol{\xi}^{\dagger}\mathrm{W}^2\boldsymbol{\xi}.\label{Q2}
\end{equation}
 The delta-function term in Eq.~(\ref{CQnormal}) represents the autocorrelation of individual atoms (shot noise), and a completely uncorrelated system is one in which  $C_{w}(\mathbf{r})= \overline{w^2}\,\delta(\mathbf{r})$. The normally ordered term in Eq.~(\ref{CQnormal}) thus represents the  correlations arising from quantum degeneracy and interaction effects.

\subsection{Static structure factor}
The $w$ static structure factor is defined as
\begin{align}
S_{w}(\mathbf{k})&\equiv \frac{1}{N}\int d\mathbf{x}\,  d\mathbf{x}'\,C_w(\mathbf{x}-\mathbf{x}')e^{-i\mathbf{k}(\mathbf{x}-\mathbf{x}')},\label{eq:SQdef0}\\
&=\frac{\langle\delta\hat{w}_{\mathbf{k}}\delta\hat{w}_{-\mathbf{k}}\rangle}{N}.\label{eq:SQdef}
\end{align}
Here $\delta\hat{w}_{\mathbf{k}}$ is the Fourier transformed fluctuation operator
\begin{eqnarray}
\delta\hat{w}_{\mathbf{k}} & \equiv & \int d\mathbf{x}\, e^{-i\mathbf{k}\cdot\mathbf{x}}\delta\hat{w}(\mathbf{x}),\label{eq:dQk_def1}\\
 & \approx & \sqrt{N}\sum_{\nu}\left(\delta\tilde{w}_{\mathbf{k}\nu}\hat{\alpha}_{\mathbf{k}\nu}+\delta\tilde{w}^*_{\mathbf{k}\nu}\hat{\alpha}_{-\mathbf{k}\nu}^{\dagger}\right),\label{eq:dQk_def2}
\end{eqnarray}
where 
\begin{equation}
\delta\tilde{w}_{\mathbf{k}\nu}\equiv\boldsymbol{\xi}^{\dagger}\mathrm{W}\mathbf{u}_{\mathbf{k}\nu}-\mathbf{v}_{\mathbf{k}\nu}^{T}\mathrm{W}\boldsymbol{\xi},\label{eq:dQKtilde_def}
\end{equation}
is a quantity we refer to as the $w$ fluctuation amplitude. In obtaining Eq.~(\ref{eq:dQk_def2}) we have neglected higher order terms in the quasiparticle operators, which should be a good approximation at low temperatures.

The static structure factor is then given by
\begin{equation}
S_{w}(\mathbf{k})=\sum_{\nu}|\delta\tilde{w}_{\mathbf{k}\nu}|^{2}\coth\left(\frac{E_{\mathbf{k}\nu}}{2k_{B}T}\right),\label{Eq:SQ}
\end{equation}
where we have used that 
\begin{equation}
\langle\hat{\alpha}^{\dagger}{}_{\mathbf{k}\nu}\hat{\alpha}{}_{\mathbf{k}'\nu'}\rangle=\frac{\delta_{\mathbf{k}\mathbf{k}'}\delta_{\nu\nu'}}{e^{E_{\mathbf{k}\nu}/k_{B}T}-1},
\end{equation}
 with $\delta_{ab}$ the Kronecker delta. 
 
In the high $k$ limit, where the kinetic energy is large compared to the thermal and interaction energies, only the uncorrelated part of $C_w$ contributes, and from Eq.~(\ref{eq:SQdef0}) we have
 \begin{equation}
 S_w(k\to\infty)=\frac{1}{n}\overline{w^2}.\label{eq:Skinf}
 \end{equation}
We refer to this as the uncorrelated limit of the structure factor.

\section{Spectra and structure factors}\label{Sec:Res1}
In this section we consider the excitations for the phases shown in Fig.~\ref{fig:PD}, and how they manifest in the various structure factors.
To do this we specialise the general discussion of the previous section to the case of total and spin density fluctuations, adopting the notation
\begin{subequations}
\begin{align}
\hat{w}&\to\{\hat{n},\hat{f}_x,\hat{f}_y,\hat{f}_z\}, \\
\mathrm{W}&\to\{\mathbb{1},\mathrm{F}_x,\mathrm{F}_y,\mathrm{F}_z\}, \\
\delta\tilde{w}_{\mathbf{k}\nu}&\to\{\delta \tilde{n}_{\mathbf{k}\nu},\delta \tilde{f}_{x,\mathbf{k}\nu},\delta\tilde{f}_{y,\mathbf{k}\nu},\delta\tilde{f}_{z,\mathbf{k}\nu}\}, \\
S_w(\mathbf{k})&\to\{S_n(\mathbf{k}),S_x(\mathbf{k}),S_y(\mathbf{k}),S_z(\mathbf{k})\}.
\end{align}
\end{subequations}
In the next subsections we discuss the various phases and their excitation spectra and fluctuations. A key set of results of our research is the analytic expressions for $S_w(\mathbf{k})$ in the $k\to0$ and $k\to\infty$ limits, for all four phases. These results are listed systematically in Table  \ref{tab:Slimits} and have been validated against numerical calculations. 
We do not present details of the lengthy derivations here.

For the most commonly realised spinor condensates of   $^{87}$Rb and $^{23}$Na atoms, the spin dependent interaction is much smaller than the spin independent interaction (see Table 2 of Ref.~\cite{Kawaguchi2012R}).  Additionally   $^{23}$Na has $c_1>0$ (i.e.~antiferromagnetic interactions), while $^{87}$Rb has $c_1<0$  (i.e.~ferromagnetic interactions). 
Here we choose to present results using $c_0 = -250 c_1$ for BA, within the range of experimental predictions for $^{87}$Rb and using $c_0 = 50 c_1$ for other phases, within the range of experimental predictions for $^{23}$Na \cite{Kawaguchi2012R}. 
We adopt the spin healing length,\footnote{The healing lengths characterize the sizes of spatial structures comparable to the relevant interaction energy, e.g.~$\hbar^2/M\xi_s^2=|c_1|n$.} $\xi_s= {\hbar}/\sqrt{M|c_1|n}$  as a convenient length scale, noting that for our choice of parameters it is a factor of $\sqrt{50}$ or $\sqrt{250}$ larger than the density healing length $\xi_n=\hbar/\sqrt{Mc_0n}$.

\begin{figure}[!t]
\includegraphics[width=3.5in]{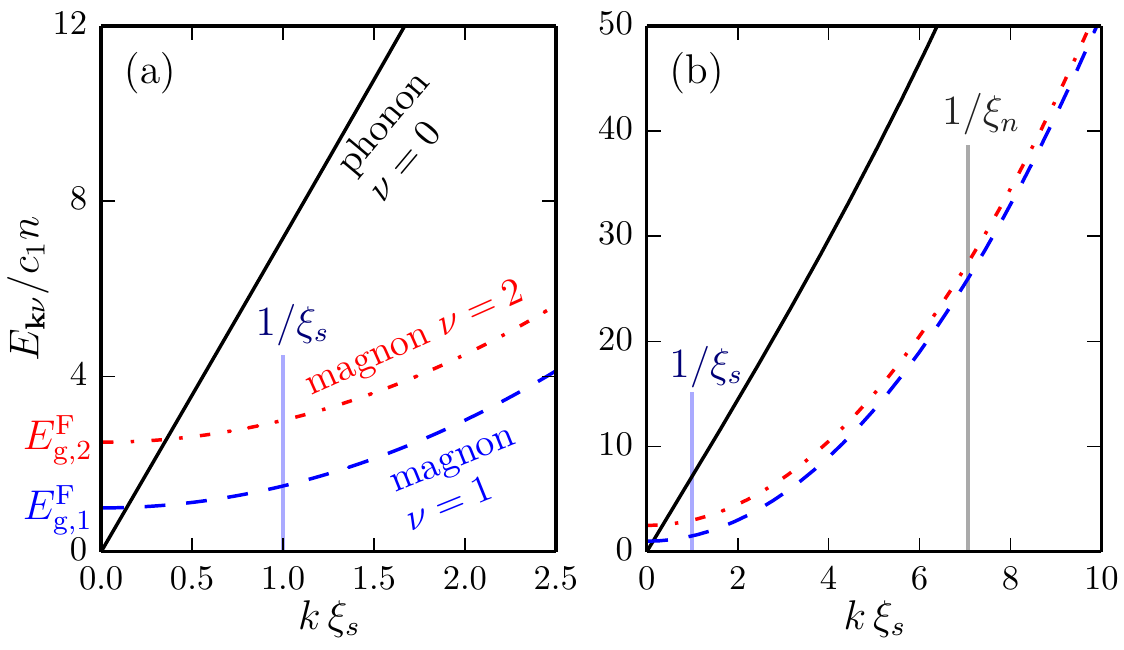}
\caption{\label{fig:Fspectrum} (Color online) Bogoliubov dispersion relations in the F
phase. Subplots (a) and (b) focus on different ranges of $k$ values. We show the phonon (solid black line), magnon  (dashed blue line),  and transverse magnon (dash-dotted red line) branches of the excitation spectra, and  attribute these the  indices $\nu=0,1,2$, respectively. 
Parameters:  $p=1.5\, c_{1}n$, $q=-c_{1}n$, $c_0=50\, c_1$, $c_1>0$.}
\end{figure}

\subsection{F phase}

\subsubsection{Condensate and excitation spectrum}
The F phase occurs for both $c_1>0$ and $c_1<0$, and in this phase the condensate is completely magnetized in the $m_F=1$ or $-1$ states depending on the value of $p$ [see Fig.~\ref{fig:PD}(a), (b)].    We focus on the case $p>0$ with atoms in the $m_F=1$ state, 
\begin{equation}
\bm{\xi}^{\mathrm{F}}=[1,0,0]^{T}.
\end{equation}
Here we have chosen $\bm{\xi}^{\mathrm{F}}$ to be real. The most general form of this state  is obtained by applying an  arbitrary gauge transformation $e^{i\chi_0}$ and a spin rotation about the  $z$-spin axis (i.e.~$e^{-i\mathrm{F}_z\chi_1}$) to $\bm{\xi}^{\mathrm{F}}$. Because the F phase is  axially symmetric  these transformations leave the properties of the condensate, and its fluctuations,  unchanged.  The nematic tensor   [see Eq.~(\ref{eq:nematic_cond})] for $\bm{\xi}^{\mathrm{F}}$ is 
\begin{equation}
q^{\mathrm{F}}= n\left(\begin{array}{ccc}
\tfrac{1}{2}  & 0 & 0\\
0 & \tfrac{1}{2}  & 0\\
0 & 0 & 1
\end{array}\right).\label{Eq:Fcondnematic}
\end{equation}

An example of the excitation spectrum for the F state \cite{Kawaguchi2012R} is shown
in Fig.~\ref{fig:Fspectrum}. This spectrum has phonon (index $\nu=0$), magnon (index $\nu=1$), and transverse magnon (index $\nu=2$) branches.\footnote{We identify the phonon branch as that making the largest contribution to the density fluctuations. For the case where the condensate has an average spin we denote the magnon modes as transverse or axial if they give rise to fluctuations that are solely transverse or solely axial to the mean spin, respectively (c.f.~\cite{Yukawa2012a}). } The phonon mode is the Nambu-Goldstone mode for this phase and resides entirely in the $m_F=1$ component. The phonon is magnetic field independent and corresponds identically to the phonon mode of a scalar gas, but with an effective interaction of  $c_0+c_1=4\pi a_2\hbar^2/M$ corresponding to the scattering length of the spin-2 channel.

The magnon modes have energy gaps  
\begin{align}
E_{\mathrm{g},1}^{\mathrm{F}}&=2p-2c_1n,\label{Eq:FEgap1}\\
E_{\mathrm{g},2}^{\mathrm{F}}&=p-q,\label{Eq:FEgap2}
\end{align}
for $\nu=1$ and $2$, respectively. 
These branches have quadratic dispersions and are magnetic field sensitive (e.g.~revealed by the dependence of $E_{\mathrm{g},1}^{\mathrm{F}}$ and $E_{\mathrm{g},2}^{\mathrm{F}}$ on $p$ and $q$).

We now consider how these modes relate to fluctuations in the system for the observables of interest. This is most easily seen by examining the fluctuation amplitudes (i.e.~$\delta\tilde{w}_{\mathbf{k}\nu}$), which reveal the contributions from the various excitation branches. By summing over these according to Eq.~(\ref{Eq:SQ}), the relevant static structure factors are then computed.

\subsubsection{Fluctuations in ${n}$ and  ${f}_{z}$ }
\begin{figure}[!t]
\includegraphics[width=3.5in]{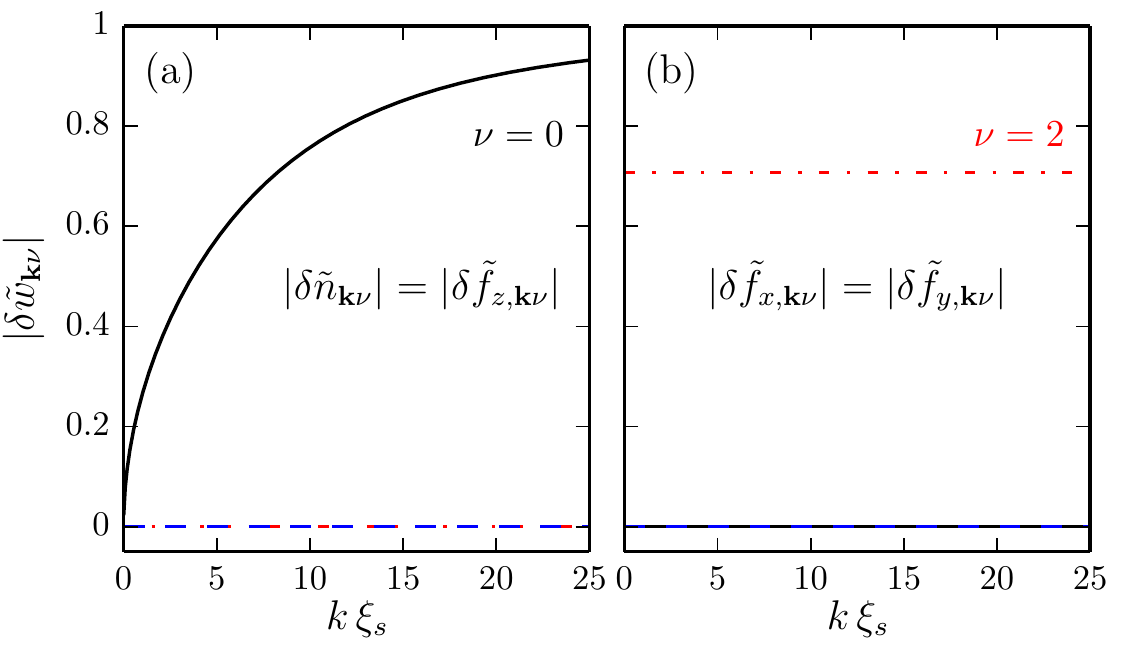}
\caption{\label{fig:Ffluctamps}(Color online)  Fluctuation amplitudes for the F
phase. Subplots (a) $\delta\tilde{n}_{\mathbf{k}\nu}$,  $\delta\tilde{f}_{z,\mathbf{k}\nu}$ and (b) $\delta\tilde{f}_{x,\mathbf{k}\nu}$,   $\delta\tilde{f}_{y,\mathbf{k}\nu}$, as defined in Eq.~(\ref{eq:dQKtilde_def}). The modes (index $\nu$) have the same line types as   in Fig.~\ref{fig:Fspectrum}. Other parameters as in Fig.~\ref{fig:Fspectrum}.} 
\end{figure}

Because the condensate resides entirely in the $m_F=1$ level we trivially have $\mathrm{F}_z\bm{\xi}^{\mathrm{F}}=\mathbb{1}\bm{\xi}^{\mathrm{F}}$ so that [from Eq.~(\ref{eq:dQKtilde_def})] the fluctuation amplitudes $\delta \tilde{n}_{\mathbf{k}\nu}$ and $\delta \tilde{f}_{z,\mathbf{k}\nu}$ are identical.\footnote{Note that for the F phase with the condensate in the $m_F=-1$ state, which we denote as $\bm{\xi}_F'$, then $\mathrm{F}_z\bm{\xi}^{\mathrm{F}'}=-\mathbb{1}\bm{\xi}^{\mathrm{F}'}$.}
The  results in Fig.~\ref{fig:Ffluctamps}(a) demonstrate that fluctuations in these quantities are entirely due to the phonon mode, with no contribution from either of the magnon modes.

\begin{figure}[!t]
\includegraphics[width=3.5in]{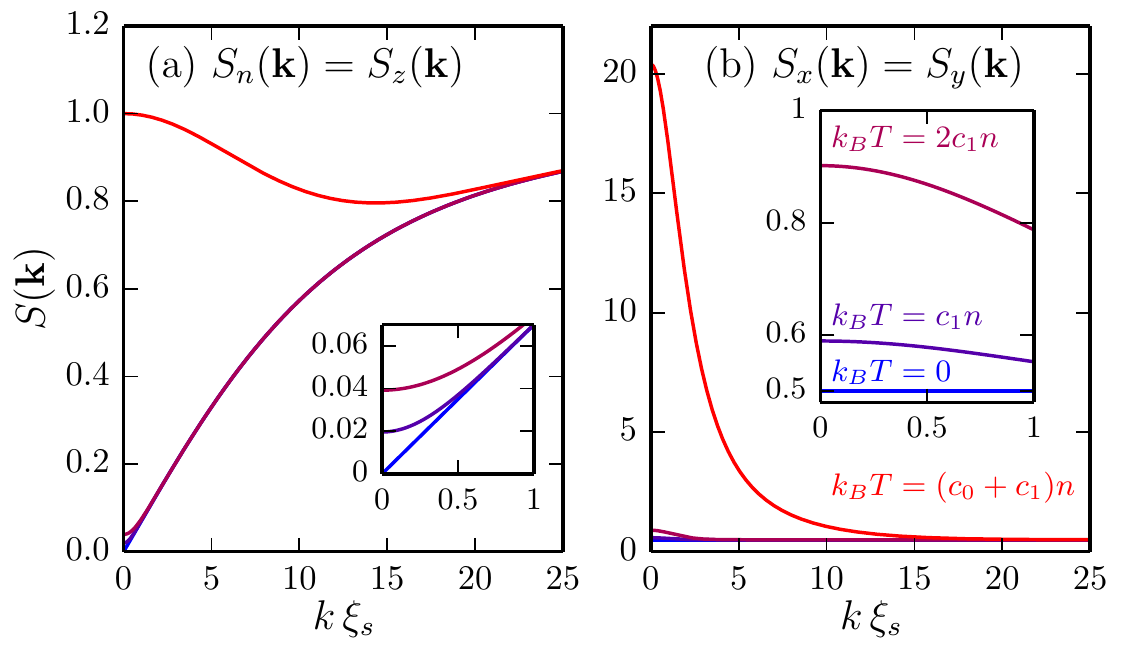}
\caption{\label{fig:FS}(Color online)  Static structure factors for the F
phase at various temperatures. Structure factors (a) $ {S_n} $ and $S_z $ (which are identical) and (b) $ S_x $ and $S_y$ (which are also identical), as defined in Eq.~(\ref{Eq:SQ}). For temperatures of (from bottom to top curves) $T=\{0,c_1,2c_1,c_0+c_1\}\!\times\!n/k_B$, as labelled in the inset to (b). Insets reveal additional detail for the lower temperature results at small $k\xi_s$. Other parameters as in Fig.~\ref{fig:Fspectrum}.}
\end{figure}

The (identical) static structure factors for density and axial spin are shown in Fig.~\ref{fig:FS}(a) for several temperatures,  with analytic limiting results given in Table~\ref{tab:Slimits}. This behavior is similar to that of the density static structure factor for a scalar Bose gas, with the phonon speed of sound set by the scattering length of the spin-2 channel ($c_0+c_1$). For example,   $S_n(\mathbf{0})=k_BT/[(c_0+c_1)n]$, and the uncorrelated limit   [$S_n(\bk)\to1$] occurs for wavevectors $k\gg1/\xi_n$ at sufficiently low temperatures  (also see Table \ref{tab:Slimits}).

\subsubsection{Fluctuations in ${f}_{x}$ and ${f}_{y}$}
The symmetry of the F phase about the spin $z$ axis is reflected in the identical fluctuations of  $f_{x}$ and $f_{y}$. Only the transverse magnon mode contributes to the fluctuation amplitudes $\delta\tilde{f}_{x,y}$, as shown in Fig.~\ref{fig:Ffluctamps}(b).   Because this   mode is single particle like (i.e.~$\mathbf{u}_{\mathbf{k}2}^T=[0,1,0]$, $\mathbf{v}_{\mathbf{k}2}^T=\mathbf{0}$), the fluctuation amplitudes are constant valued with $|\delta \tilde{f}_{x,\mathbf{k}2}|=|\delta \tilde{f}_{x,\mathbf{k}2}|=1/\sqrt{2}$. 
Note that the $\nu=1$ magnon mode is of the form $\mathbf{u}_{\mathbf{k}1}^T=[0,0,1]$, $\mathbf{v}_{\mathbf{k}1}^T=\mathbf{0}$, and  does not contribute to total density or spin density fluctuations.

The associated structure factors, $S_x$ and $S_y$, are shown in Fig.~\ref{fig:FS}(b), with analytic limiting results given in Table~\ref{tab:Slimits}.   These  factors have a non-zero value for $k\to0$  at $T=0$, i.e.~$S_{x,y}^{T=0}(k\to0)\ne0$.
This behaviour was also found for a two-component system in Ref.~\cite{Abad2013a}, where the magnon mode was also energetically gapped.  The energy gap of the transverse magnon mode delays the onset of thermal fluctuations to temperatures $T\gtrsim E_{\mathrm{g},2}^{\mathrm{F}}/k_B$.

\subsection{P phase}

\subsubsection{Condensate and excitation spectrum}
The  P  phase occurs for both $c_1>0$ and $c_1<0$ [see Figs.~\ref{fig:PD}(a),(b)]. In this phase the condensate is  unmagnetized  and occupies the $m_F=0$ level, with normalised spinor
\begin{equation}
\bm{\xi}^{\mathrm{P}}=[0,1,0]^{T}.
\end{equation}
The nematic tensor   [see Eq.~(\ref{eq:nematic_cond})] for $\bm{\xi}^{\mathrm{P}}$ is 
\begin{equation}
q^{\mathrm{P}}= n\left(\begin{array}{ccc}
1 & 0 & 0\\
0 & 1  & 0\\
0 & 0 & 0
\end{array}\right).\label{Eq:Pcondnematic}
\end{equation}
The most general form of the P-phase spinor is obtained by applying an arbitrary gauge transformation  and a spin rotation about the  $z$-spin axis to $\bm{\xi}^{\mathrm{P}}$. Because the P phase is  axially symmetric  these transformations leave the properties of the condensate, and its fluctuations, unchanged.

\begin{figure}[!t]
\includegraphics[width=3.5in]{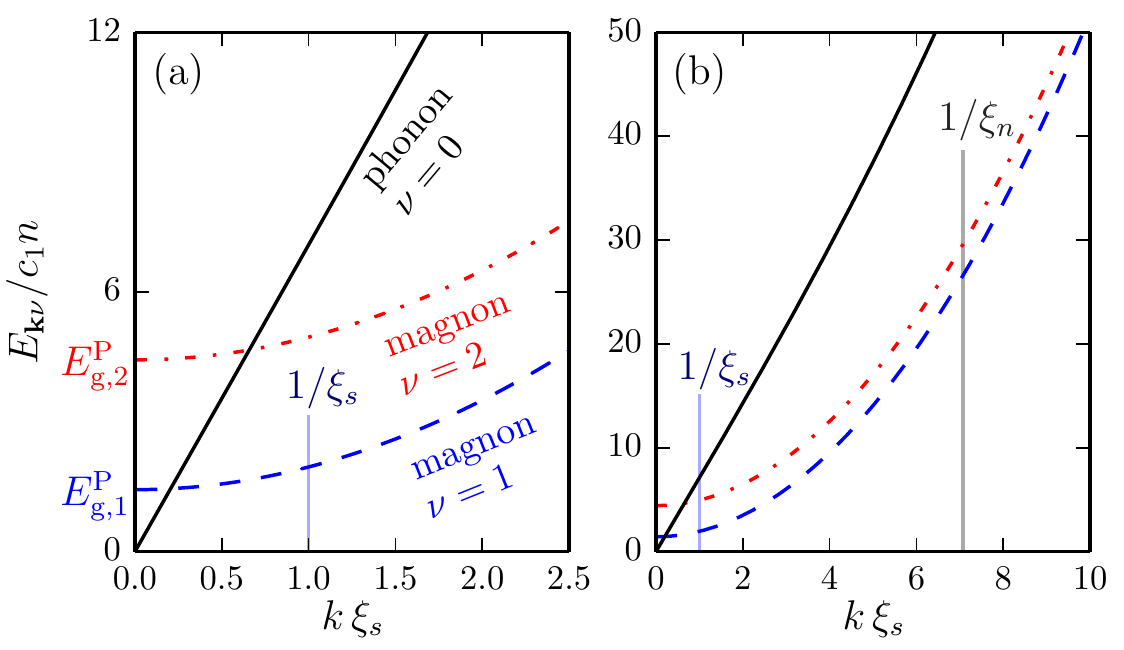}
\caption{\label{fig:Pspectrum} (Color online) Bogoliubov dispersion relations in the P
phase. Subplots (a) and (b) focus on different ranges of $k$ values. We show the branches of the excitation spectra for the phonon mode (solid black line) and the two magnon modes (dashed blue line and dash-dotted red line). 
Parameters: $q=2.1\,c_1n$, $c_0=50\, c_1$ and $p=1.5\, c_1n$, $c_1>0$.}
\end{figure}

An example of the excitation spectrum for the P state \cite{Kawaguchi2012R} is shown
in Fig.~\ref{fig:Pspectrum}. This spectrum is similar to the F phase [Fig.~\ref{fig:Fspectrum}] in that it has a phonon (index $\nu=0$) and two gapped magnon branches (indices $\nu=1,2$). 
 The magnon gaps depend on the magnetic field and are given by
\begin{align}
E_{\mathrm{g},1}^{\mathrm{P}}&=\sqrt{q(q+2c_1n)}-p,\label{Eq:PEgap1}\\
E_{\mathrm{g},2}^{\mathrm{P}}&=\sqrt{q(q+2c_1n)}+p,\label{Eq:PEgap2}
\end{align}
for $\nu=1$ and $2$, respectively.
The $\nu=1$ magnon mode is of the form $\mathbf{u}_{\mathbf{k}1}^T=[u,0,0]$, $\mathbf{v}_{\mathbf{k}1}^T=[0,0,v]$, while the $\nu=2$ magnon mode has $\mathbf{u}_{\mathbf{k}2}^T=[0,0,u]$, $\mathbf{v}_{\mathbf{k}2}^T=[v,0,0]$. The phonon mode resides entirely in the $m_F=0$ component and corresponds identically to that of a scalar gas  with an effective interaction of  $c_0$.

\subsubsection{Fluctuations in ${n}$ and  ${f}_{z}$ }
Because the condensate resides entirely in the $m_F=0$ level we have that the $f_z$ fluctuations are identically zero [from Eq.~(\ref{eq:dQKtilde_def})] to the level of approximation we work at here, with the leading order term coming from the small terms we neglected in Eq.~(\ref{eq:dQk_def2}). We do not consider a higher order treatment here, and take the $f_z$ fluctuations to be zero.

The density fluctuations are entirely due to the phonon mode, which resides in $m_F=0$, with no contribution from either of the magnon modes [see     
Fig.~\ref{fig:Pfluctamps}(a)] . The associated  static structure factor is shown in Fig.~\ref{fig:PS}(a) for several temperatures, with analytic limiting results given in Table~\ref{tab:Slimits}.

\subsubsection{Fluctuations in ${f}_{x}$ and ${f}_{y}$}
\begin{figure}[!t]
\includegraphics[width=3.5in]{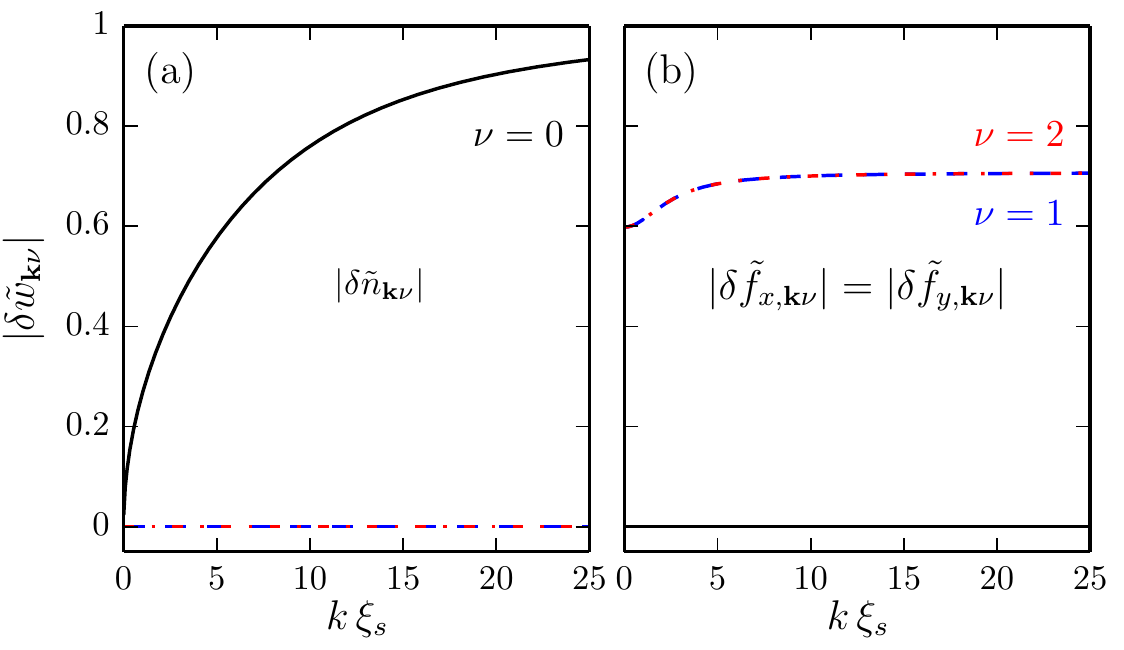}
\caption{\label{fig:Pfluctamps}(Color online)  Fluctuation amplitudes for the P
phase. Subplots (a) $\delta\tilde{n}_{\mathbf{k}\nu}$,  and (b) $\delta\tilde{f}_{x,\mathbf{k}\nu}$,  $\delta\tilde{f}_{y,\mathbf{k}\nu}$, as defined in Eq.~(\ref{eq:dQKtilde_def}). Note $\delta\tilde{f}_{z,\mathbf{k}\nu}=0$. The modes (index $\nu$) have the same line types as   in Fig.~\ref{fig:Pspectrum}. Other parameters as in Fig.~\ref{fig:Pspectrum}.} 
\end{figure}

\begin{figure}[!t]
\includegraphics[width=3.5in]{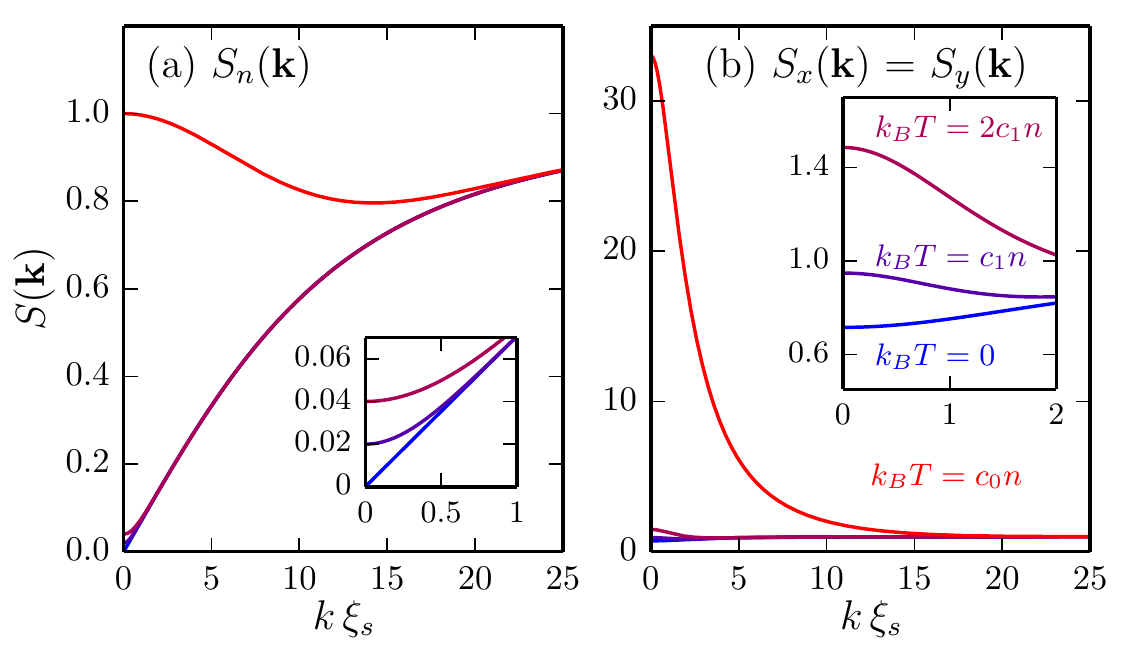}
\caption{\label{fig:PS}(Color online)  Static structure factors for the P
phase at various temperatures. Structure factors  (a) $ {S_n} $ and (b) $ S_x $, $ S_y $, as defined in Eq.~(\ref{Eq:SQ}). Note $S_z=0$. For temperatures of (from bottom to top curves) $T=\{0,c_1,2c_1,c_0\}\!\times\!n/k_B$, as labelled in the inset to (b). Insets reveal additional detail for small $k\xi_s$. Other parameters as in Fig.~\ref{fig:Pspectrum}.} 
\end{figure}

Because the P phase is axisymmetric, the $f_{x}$ and $f_{y}$ fluctuations are identical, and relevant fluctuation amplitudes are shown in Fig.~\ref{fig:Pfluctamps}(b).  These results show that both magnon modes contribute equally. The associated structure factors are shown in Fig.~\ref{fig:PS}(b), with analytic limiting results given in Table~\ref{tab:Slimits}.
Similarly to the  $S_x$ and $S_y$ structure factors considered for the F phase, these  are also gapped at $k\to0$ and at zero temperature.

\subsection{AF phase}

\subsubsection{Condensate and excitation spectrum}
The AF phase  occurs only for $c_1>0$ [see Fig.~\ref{fig:PD}(a)]. In this phase the condensate
takes the form 
\begin{equation}
\bm{\xi}^{\mathrm{AF}}=\left[\sqrt{\tfrac{1}{2}(1+{f}_z/n)},0,\sqrt{\tfrac{1}{2}(1-f_z/n)}\right]^T,
\end{equation}
and has a $z$-component of magnetization given by $f_z =p/c_1$ for  
$|p|\le c_1n$.  The AF state breaks symmetry about the $z$ axis, as can be seen from its  nematic tensor  [see Eq.~(\ref{eq:nematic_cond})],
\begin{equation}
q^{\mathrm{AF}}= n\left(\begin{array}{ccc}
\tfrac{1}{2}(1+\alpha_z) & 0 & 0\\
0 & \tfrac{1}{2}(1-\alpha_z) & 0\\
0 & 0 & 1
\end{array}\right),\label{Eq:AFcondnematic}
\end{equation}
where we have introduced the variable 
\begin{equation}
\alpha_z=\sqrt{1-(f_z/n)^2}.
\end{equation}

\begin{figure}[!t]
\includegraphics[width=3.5in]{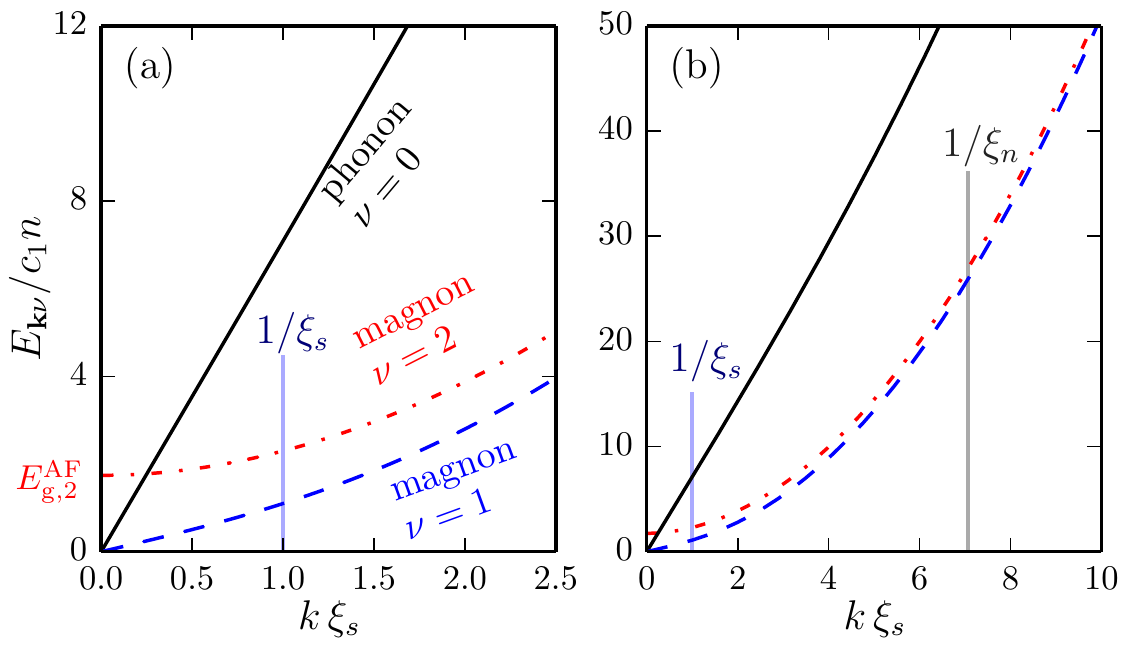}
\caption{\label{fig:AFspectrum} (Color online) Bogoliubov dispersion relations in the AF phase. Subplots (a) and (b) focus on different ranges of $k$ values. We show the phonon (solid black line), axial magnon  (dashed blue line),  and transverse magnon (dash-dotted red line) branches of the excitation spectra and  attribute these the  indices $\nu=0,1,2$, respectively. 
Parameters: $q=-c_1n$, $c_0=50\, c_1$, and $p=0.2\, c_1n$, giving $f_z=0.2\,n$, $\alpha_z\approx0.98$.}
\end{figure}

\begin{figure}[!t]
\includegraphics[width=3.5in]{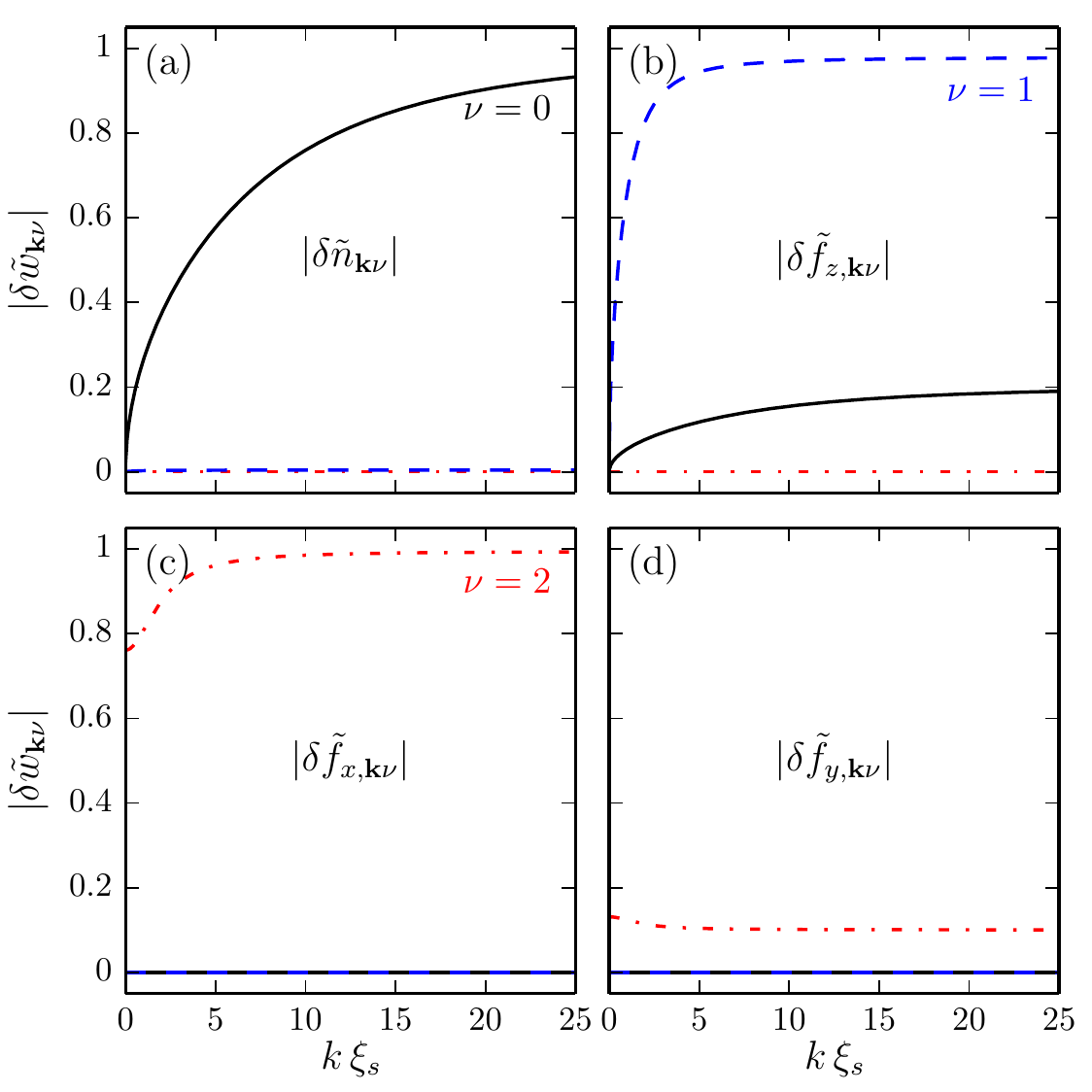}
\caption{\label{fig:AFfluctamps}(Color online)  Fluctuation amplitudes for the AF phase. Subplots (a) $\delta\tilde{n}_{\mathbf{k}\nu}$, (b) $\delta\tilde{f}_{z,\mathbf{k}\nu}$, (c) $\delta\tilde{f}_{x,\mathbf{k}\nu}$, and (d) $\delta\tilde{f}_{y,\mathbf{k}\nu}$, as defined in Eq.~(\ref{eq:dQKtilde_def}). The modes (index $\nu$) have the same line types as   in Fig.~\ref{fig:AFspectrum}. Other parameters as in Fig.~\ref{fig:AFspectrum}.} 
\end{figure}

\begin{figure}[!t]
\includegraphics[width=3.5in]{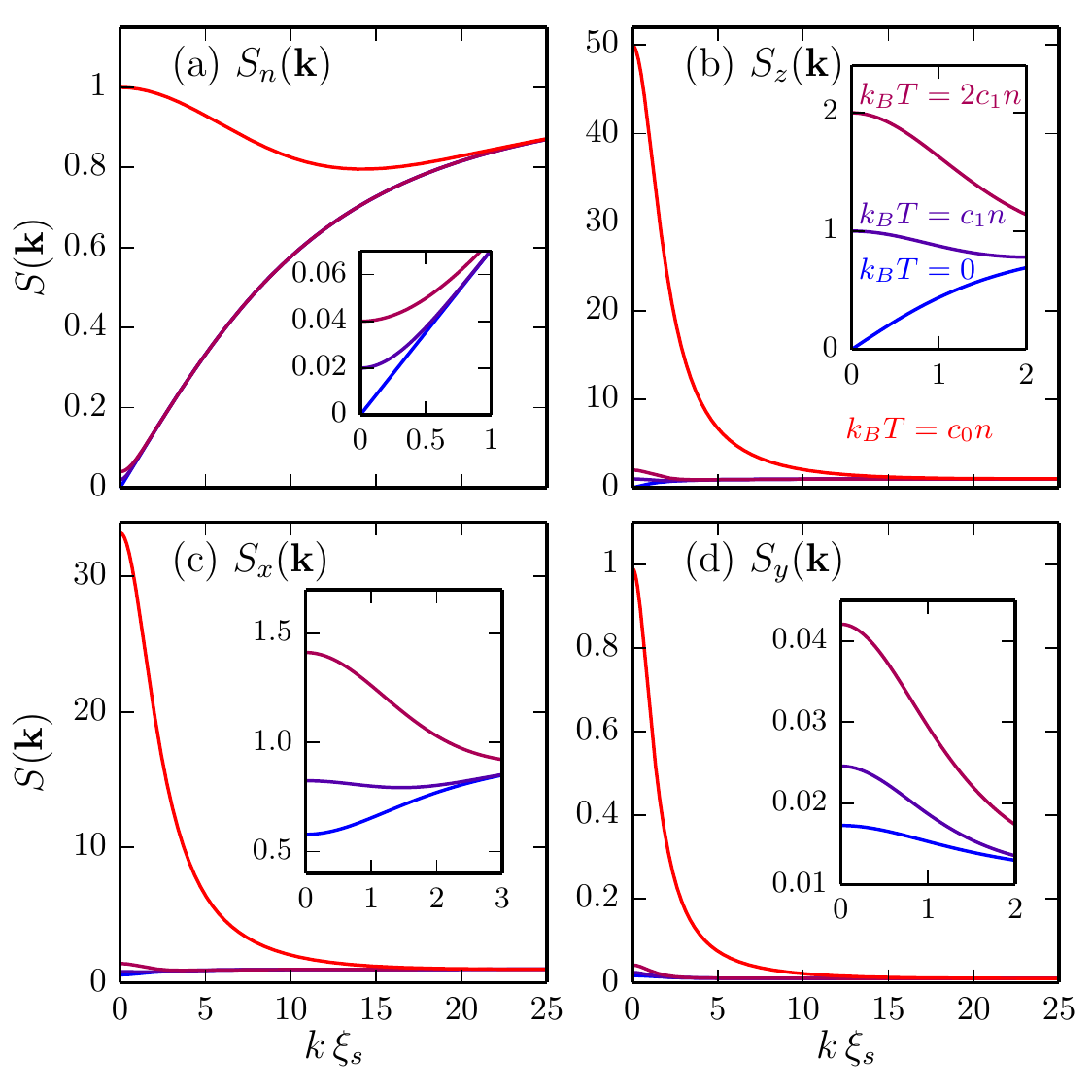}
\caption{\label{fig:AFS}(Color online)  Static structure factors for the AF phase at various temperatures. Structure factors (a) $ {S_n} $, (b) $S_z$, (c) $ S_x $, and (d) $ S_y$, as defined in Eq.~(\ref{Eq:SQ}). For temperatures of (from bottom to top curves) $T=\{0,c_1,2c_1,c_0\}\!\times\!n/k_B$, as labelled in the inset to (b). Insets reveal the lower temperature results at small $k\xi_s$. Other parameters as in Fig.~\ref{fig:AFspectrum}.} 
\end{figure}

We note that $q^{\mathrm{AF}}$ corresponds to $q^{\mathrm{F}}$ (\ref{Eq:Fcondnematic}) in the limit of a fully magnetized AF state (i.e.~$f_z\to n$).
The most general form of the AF phase spinor is obtained by applying  an arbitrary gauge transformation   and a spin rotation about the $z$-spin axis to $\bm{\xi}^{\mathrm{AF}}$.   We note that the spin rotation changes the orientation of the nematic distortion in the spin-$xy$ plane.

An example of the AF excitation spectrum \cite{Kawaguchi2012R} is shown
in Fig.~\ref{fig:AFspectrum}. 
It has phonon (index $\nu=0$), axial magnon (index $\nu=1$), and transverse magnon (index $\nu=2$) branches.  The AF phase has two broken continuous symmetries giving rise to two Nambu-Goldstone modes: in addition to the phonon mode arising from the broken $U(1)$ symmetry of the condensate, the broken axial spin symmetry (revealed by the nematic tensor) yields a massless axial magnon mode.
The axial magnon dispersion mode crosses over from having a linear to quadratic dependence on $k$ 
at a wavevector of $k\sim1/\xi_{s}$, whereas the phonon mode crosses
over at $k\sim1/\xi_{n}$. 
The transverse magnon mode has an energy gap of 
\begin{equation}
E_{\mathrm{g},2}^{\mathrm{AF}}=c_1n\sqrt{(1-q/c_1n)^{2}-\alpha^{2}_z}.\label{Eq:AFEgap}
\end{equation}
 
\subsubsection{Fluctuations in ${n}$}
The density fluctuation amplitudes $\delta \tilde{n}_{\mathbf{k}\nu}$ are shown in Fig.~\ref{fig:AFfluctamps}(a). These results demonstrate that the density fluctuations are dominated by the phonon mode, although a weak contribution arises from the axial magnon mode. This magnon contribution increases as $c_1$  increases relative to $c_0$  and also depends on the axial magnetization $f_z$. (Note: The axial magnon and phonon modes decouple for $f_z=0$, and at this point the magnon mode does not contribute to $\delta \tilde{n}_{\mathbf{k}\nu}$.)

The density static structure factor ($S_n$) is shown in Fig.~\ref{fig:AFS}(a) for several temperatures. This behavior is similar to that of the density static structure factor in the F phase, except that the phonon speed of sound varies between the value set by $c_0n$ and $(c_0+c_1)n$, depending on $f_z$.

\subsubsection{Fluctuations in $f_{z}$}
The axial spin   fluctuation amplitudes   $\delta \tilde{f}_{z,\mathbf{k}\nu}$ are shown in Fig.~\ref{fig:AFfluctamps}(b), and demonstrate a dominant contribution from the axial magnon mode, and a smaller, but appreciable contribution from the phonon mode. The associated static structure factor ($S_z$) is shown in Fig.~\ref{fig:AFS}(b). The general behavior is similar to the density fluctuation case, but with the much smaller spin-dependent energy $c_1n$ being the appropriate energy scale. Thus the fluctuations are  more easily thermally activated and the uncorrelated limit [$S_z(\bk)\to1$] is reached at lower wave vectors $k\gg1/\xi_s$ [also see Table \ref{tab:Slimits}].
 
\subsubsection{Fluctuations in ${f}_{x}$ and ${f}_{y}$}
The transverse spin fluctuation amplitudes, i.e.~$\delta \tilde{f}_{x,\mathbf{k}\nu}$  and  $\delta \tilde{f}_{y,\mathbf{k}\nu}$, are shown in Figs.~\ref{fig:AFfluctamps}(c) and (d), respectively.  Only the transverse magnon mode contributes to these. The difference in the behavior of $\delta \tilde{f}_{x,\mathbf{k}\nu}$ and  $\delta \tilde{f}_{y,\mathbf{k}\nu}$ reveals the broken symmetry of the AF state about the $z$-spin axis [c.f.~Eq.~(\ref{Eq:AFcondnematic})].

The associated structure factors are shown in Figs.~\ref{fig:AFS}(c) and (d). Similarly to the  $S_x$ and $S_y$ structure factors  for the F and P phases, these also have a non-zero value for $k\to0$  at $T=0$. The energy gap of the transverse magnon mode delays the onset of thermal fluctuations to temperatures $T\gtrsim E_{\mathrm{g},2}^{\mathrm{AF}}/k_B$.

\subsection{BA phase}

\subsubsection{Condensate and excitation spectrum}
The BA phase occurs for $c_1<0$ [see Fig.~\ref{fig:PD}(b)], and in this phase the condensate occupies all three  $m_F$ states.
The results we present here are for the case of $p=0$, where the magnetization is purely transverse (i.e.~$f_z=0$). This case has the advantage that it affords a simpler analytic treatment \cite{Murata2007a,Uchino2010a}, allowing us to write the spinor as
\begin{equation}
\bm{\xi}^{\mathrm{BA}}=\left[\tfrac{1}{2}\sqrt{1 - \tilde{q}},\sqrt{\tfrac{1}{2}(1 + \tilde{q})}, \tfrac{1}{2}\sqrt{1 - \tilde{q}}\right]^T,
\end{equation}
where $\tilde{q} \equiv q/2|c_1| n$.
For our choice of a real spinor $\bm{\xi}^{\mathrm{BA}}$, the magnetization is along the $x$-spin axis. The most general form of the BA phase spinor is given by  an arbitrary gauge transformation   and  spin rotation about the $z$-spin axis applied to $\bm{\xi}^{\mathrm{BA}}$.
The nematic tensor for $\bm{\xi}^{\mathrm{BA}}$ [see Eq.~(\ref{eq:nematic_cond})] is
\begin{equation}
q^{\mathrm{BA}}= n\left(\begin{array}{ccc}
1 & 0 & 0\\
0 & \frac{1}{2}(1 + \tilde{q}) & 0\\
0 & 0 & \frac{1}{2}(1 - \tilde{q})
\end{array}\right),\label{Eq:BAcondnematic}
\end{equation}
which reveals the broken symmetry of the BA state about the $z$ axis due to the transverse magnetization $f_x = \sqrt{1-\tilde{q}^2}$.

\begin{figure}[!t]
\includegraphics[width=3.5in]{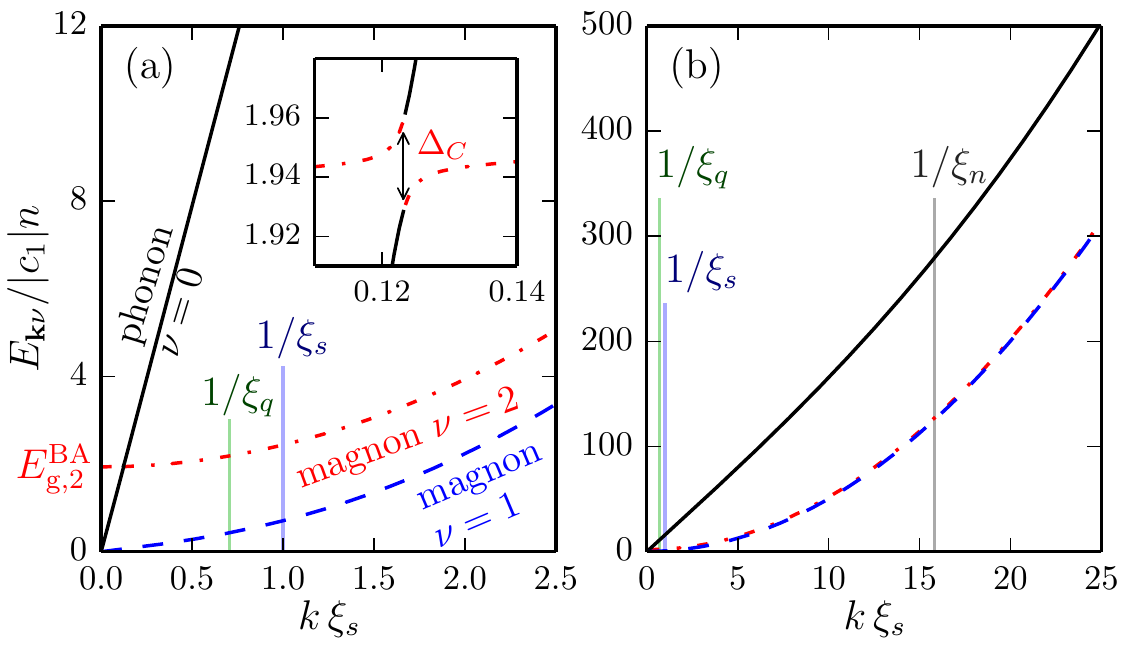}
\caption{\label{fig:BAspectrum} (Color online) Bogoliubov dispersion relations in the BA
phase. Subplots (a) and (b) focus on different ranges of $k$ values. We show the branches of the excitation spectra for the phonon mode (solid black line) and the axial (dashed blue line; note the magnetization axis is $x$) and transverse (dash-dotted red line) magnon modes. Inset to (a) reveals an avoided crossing between the phonon and transverse magnon modes.
Parameters: $q=0.5\,|c_1|n$, $c_0=-250\, c_1$, $p=0$ and $c_1<0$.}
\end{figure}

The Bogoliubov excitations of the BA phase have been investigated in several recent papers \cite{Murata2007a,Uchino2010a} (also see Appendix of \cite{Barnett2011a}).\footnote{While various analytic results have been reported for the $p=0$ case \cite{Uchino2010a}, the understanding of the $p\ne0$ case is based largely on numerical results \cite{Murata2007a}. }
The excitations of the BA phase at $p=0$ are shown in Fig.~\ref{fig:BAspectrum}.  Because the BA phase has two broken continuous symmetries, the system has two gapless Nambu-Goldstone modes:  a phonon branch (index $\nu=0$) and a transverse magnon branch (index $\nu=1$). These two modes are decoupled at $p=0$.
The transverse magnon has the energy dispersion $E_{\mathbf{k}1}=\sqrt{\epsilon_{\mathbf{k}}(\epsilon_{\mathbf{k}}+q)}$ \cite{Uchino2010a}, i.e.~independent of the interaction parameters, with $\efree \equiv \hbar^2k^2/2M$ the free particle energy. Since the quadratic Zeeman sets the relevant energy scale for this magnon, we define an associated length scale $\xi_q\equiv\hbar/\sqrt{Mq}$.
The last branch (index $\nu=2$) is a magnon excitation with energy gap 
\begin{equation}
E^{\mathrm{BA}}_{\mathrm{g},2}=2|c_1|n\sqrt{1-\tilde{q}^2}=2f_x|c_1|n.
\end{equation}
This gapped magnon does couple to the phonon branch, and   they have an avoided crossing, as revealed in Fig.~\ref{fig:BAspectrum}(a) and inset. We have chosen to switch the labelling on either side of this crossing to match the labelling choice made in Ref.~\cite{Murata2007a} and also to ensure that away from the crossing the $\nu=0$ mode has a phonon character (i.e.~a dominant effect on density fluctuations). The coupling between these two modes is small so that the avoided crossing occurs over a narrow range of $k$ vectors, with the energy gap between these branches being
\begin{equation}
\Delta_C =q\sqrt{1-\tilde{q}^2}\sqrt{\frac{|c_1|}{c_0}},
\end{equation}
to lowest order in $c_1/c_0$.

\begin{figure}[!t]
\includegraphics[width=3.5in]{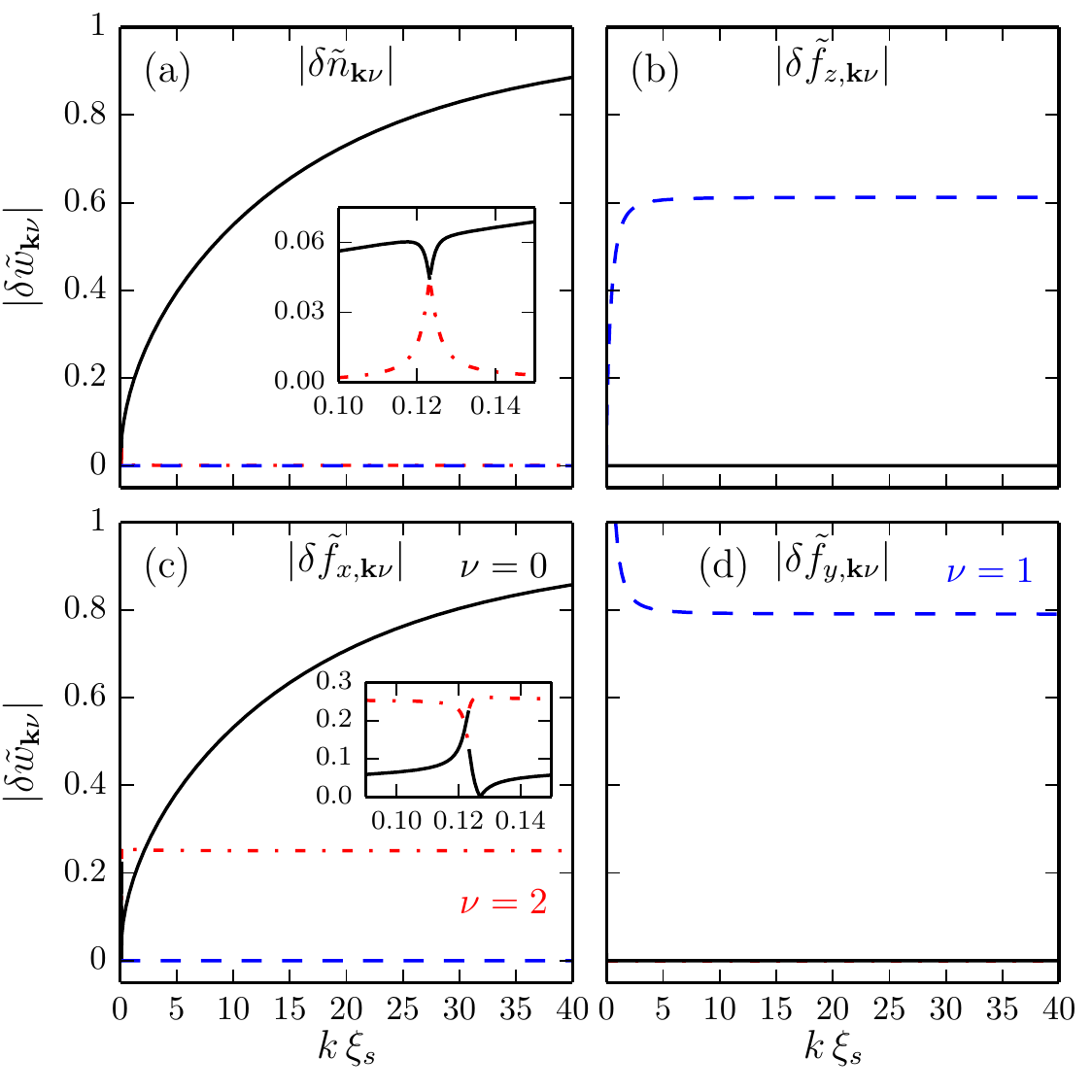}
\caption{\label{fig:BAfluctamps}(Color online)  Fluctuation amplitudes for the BA
phase. Subplots (a) $\delta\tilde{n}_{\mathbf{k}\nu}$, (b) $\delta\tilde{f}_{z,\mathbf{k}\nu}$, (c) $\delta\tilde{f}_{x,\mathbf{k}\nu}$, and (d) $\delta\tilde{f}_{y,\mathbf{k}\nu}$, as defined in Eq.~(\ref{eq:dQKtilde_def}). The modes (index $\nu$) have the same line types as   in Fig.~\ref{fig:BAspectrum}. Insets to (a) and (c) reveal additional detail for $k\xi_s$ close to the avoided crossing. Other parameters as in Fig.~\ref{fig:BAspectrum}.} 
\end{figure}

\begin{figure}[!h]
\includegraphics[width=3.5in]{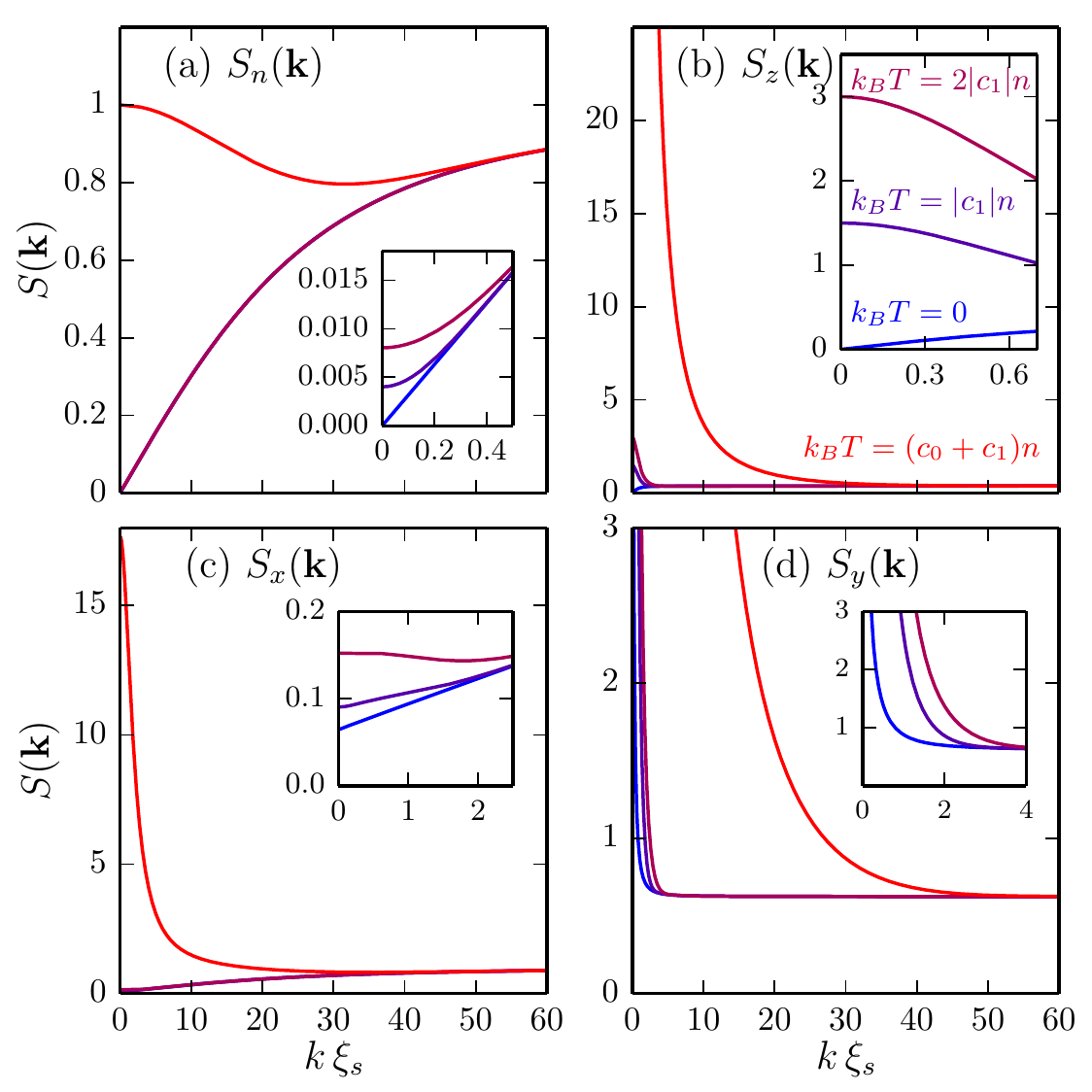}
\caption{\label{fig:BAS}(Color online)  Static structure factors for the BA
phase at various temperatures. Structure factors  (a) $ {S_n} $, (b) $S_z $, (c) $ S_x $, and (d) $ S_y $, as defined in Eq.~(\ref{Eq:SQ}). For temperatures of (from bottom to top curves) $T=\{0,|c_1|,2|c_1|,c_0+c_1\}\!\times\!n/k_B$, as labelled in the inset to (b). Insets reveal additional detail for small $k\xi_s$. Other parameters as in Fig.~\ref{fig:BAspectrum}.} 
\end{figure}

\subsubsection{Fluctuations in $n$ and $f_z$}
From Fig.~\ref{fig:BAfluctamps}(a) we see that the dominant contribution to density fluctuations comes from the phonon mode, although a contribution from the (gapped) axial magnon mode occurs near the avoided crossing noted in the spectrum.\footnote{The rapid variation in $\{\delta\tilde{n}_{\mathbf{k}0},\delta\tilde{n}_{\mathbf{k}2}\}$ and $\{\delta\tilde{f}_{x,\mathbf{k}0},\delta\tilde{f}_{x,\mathbf{k}2}\}$ for $k\approx0.12/\xi_s$ occurs because the phonon and transverse magnon hybridize near the anti-crossing. We emphasise that the summed contribution of these fluctuations to the relevant structure factors is smooth.} In contrast, the $f_z$ fluctuations  come entirely from the (Nambu-Goldstone) transverse magnon branch ($\nu=1$) [see Fig.~\ref{fig:BAfluctamps}(b)]. 

The structure factors $S_n$ and $S_z$ are shown in Fig.~\ref{fig:BAS}(a) and (b), respectively, with analytic expressions for the limiting behavior provided in Table~\ref{tab:Slimits}. Interestingly, the long wavelength fluctuations of the $z$ component of magnetization is set by the quadratic Zeeman energy, i.e.~$S_z(0)=(1 - \tilde{q}) k_BT/q$. This diverges for $q\to0$  as the full spin rotational symmetry [$SO(3)$] is restored (noting that we have set $p=0$).

\setlength{\extrarowheight}{9pt}

  \begin{table*}[!htbp]
   \centering
   {\small 
   \begin{tabular}{c|c|C|C|C } 
       \hline
       Phase &  Observable(s) & \multicolumn{2}{C|}{S_w(k\to0)} & S_w(k\to\infty)   \\\cline{3-4}
       & & T=0 & T>0 & \\\hline
   F &     ${n},{f}_z$ & \sqrtf{\efree}{2(c_0+c_1)n}  &  \frac{k_BT}{ (c_0+c_1)n}  & 1\\\cline{2-5}
    &    ${f}_x,{f}_y$ & \multicolumn{2}{C|}{\frac{1}{2}\coth\left(\frac{E_{\mathrm{g},2}^{\mathrm{F}}}{2k_BT}\right)}   & \frac{1}{2}\\\hline
    P &   ${n}$ & \frac{k\xi_n}{2}        &  \frac{k_BT}{c_0n}   & 1\\\cline{2-5}
    &   $ {f}_z$ & \multicolumn{2}{C|}{ 0} & 0 \\\cline{2-5}
     &   ${f}_x,{f}_y$ & \multicolumn{2}{C|}{\frac{\coth\left(\frac{E_{\mathrm{g},1}^{\mathrm{P}}}{2k_BT}\right)+\coth\left(\frac{E_{\mathrm{g},2}^{\mathrm{P}}}{2k_BT}\right)}{2\sqrt{1+2c_1n/q}}} & 1\\\hline
       AF &      $n$ &  \left[1+\frac{f_z^2}{n^2}\left(-\frac{1}{2}\frac{c_1}{c_0}  + \sqrt{1-\frac{f_z^2}{n^2}}\frac{c_1^{3/2}}{c_0^{3/2}} 
  \right)\right] \frac{k\xi_n}{2}  & \frac{k_BT}{c_0n} &  1\\\cline{2-5}   
   &   ${f}_z$ & \left[\sqrt{1-\frac{f_z^2}{n^2}}\left(1-\frac{3f_z^2}{2n^2}\frac{c_1}{c_0} \right) +  \frac{f_z^2}{n^2}\sqrt{\frac{c_1}{c_0}}  
    \right] \frac{k\xi_s}{2} & \frac{k_BT}{c_1n}  & 1    \\\cline{2-5}
    &  $\genfrac{}{}{0pt}{}{ f_x \to + }{f_y \to - }$ & \multicolumn{2}{C|}{\frac{1}{2}\left(\frac{1\pm\alpha_z}{1\pm\alpha_z-q/c_1 n} \right) \frac{E_{\mathrm{g},2}^{\mathrm{AF}}}{c_1 n}\coth\left(\frac{E_{\mathrm{g},2}^{\mathrm{AF}}}{2k_BT}\right) } 
    &  \frac{1}{2}(1\pm\alpha_z) \\\hline
     BA   &   $n$ & \sqrt{\frac{\efree}{2(c_0+c_1)n}} & \frac{k_BT}{ (c_0+c_1)n} & 1 \\\cline{2-5}
 $(p=0)$&  $f_x$  & \frac{\qt^2}{\sqrt{1-\qt^2}} +  \frac{1}{1-\qt^2}\sqrtf{\efree}{2(c_0+c_1)n} 
 & \frac{\kt}{(c_0+c_1)n}\frac{1}{1-\qt^2} +\frac{\qt^2}{\sqrt{1-\qt^2}} \coth\fracb{E_{\mathrm{g},2}^{\mathrm{BA}}}{2k_BT}  & 1 \\\cline{2-5}
     & $f_y$ & \frac{1}{2}(1 + \tilde{q})\sqrt{\frac{q}{\efree}}    & (1 + \tilde{q})\frac{k_B T}{\efree} & \frac{1}{2}(1 + \tilde{q})\\\cline{2-5}
  &    $f_z$ & \frac{1}{2}(1 - \tilde{q})\sqrt{\frac{\efree}{q}}   & (1 - \tilde{q})\frac{k_BT}{q}   & \frac{1}{2}(1 - \tilde{q})\\\hline\hline
       \end{tabular}
   \caption{Large and small $k$ limits of the structure factors. Where necessary in the $k\to0$ limits we distinguish between $T=0$ and $T>0$ results: In the  $T=0$  case we give a $k$ expansion, whereas for $T>0$ we give the structure factor value at $k=0$. For $n$ and $f_z$ in the AF phase, the $T=0$ results are the first terms in an expansion for $c_1\ll c_0$ and the finite $T$ results are valid for $|f_z|<n$.}
   \label{tab:Slimits}
   }
\end{table*}

\subsubsection{Fluctuations in $f_x$ and $f_y$}
Because the magnetization lies along $x$ for our choice of $p$ and $\bm{\xi}^{\mathrm{BA}}$, fluctuations in $f_x$ correspond to fluctuations in the length of the magnetization.
Fig.~\ref{fig:BAfluctamps}(c) reveals that both the gapped magnon mode and the phonon mode contribute to these fluctuations. In contrast, fluctuations in $f_y$ are orthogonal to the direction of magnetization and act to restore the  axial symmetry [$SO(2)$] of the Hamiltonian. In Fig.~\ref{fig:BAfluctamps}(d) we see that these fluctuations are entirely due to the (Nambu-Goldstone) transverse magnon mode, and that these fluctuations diverge as $k\to0$. 
 The divergence is clearly apparent in $S_y$ [see Fig.~\ref{fig:BAS}(d)] and is seen to go as $k^{-2}$ for small $k$ at finite temperature [see Table~\ref{tab:Slimits}].

\subsubsection{BA phase for $p\ne0$}
We conclude by briefly commenting on the qualitative behavior for $p\ne0$. In this case the condensate magnetization  tilts out of the $xy$-plane and the Nambu-Goldstone branches (i.e.~$\nu=0$ and $\nu=1$ branches) become coupled (c.f.~at $p=0$, where the only coupling is between the $\nu=0$ and $\nu=2$ branches, giving rise to the avoided crossing). In Ref.~\cite{Murata2007a} this occurrence was referred to as phonon-magnon coupling. As a result of this coupling the $\nu=1$ mode contributes to density fluctuations, and the $\nu=0$ mode contributes to $f_z$ fluctuations.  

\section{Discussion and Conclusions}\label{Sec:Discussion-Conclusion}
In this paper we have developed a formalism for the static structure factor of a uniform spin-1  condensate subject to constant linear and quadratic Zeeman shifts. Our results are based on the Bogoliubov formalism and are accurate to the leading order term proportional to the condensate density. The static structure factors are an important tool in quantifying fluctuations for scalar and binary systems (e.g.~see \cite{Astrakharchik2007a,Klawunn2011a,Recati2011a,Bisset2013}), and this work is important for extending such results to the spinor system. 

\setlength{\extrarowheight}{9pt}
 \begin{table*}[!htbp] 
   \centering
   {\small 
   \begin{tabular}{c|c|C|C|C } 
       \hline
       Phase & Observable(s) & \multicolumn{3}{C}{ |\delta \wh_{\bk\nu}|^2} \\\cline{3-5}
             &            &     \nu=0 & \nu=1 & \nu=2 \\\hline
       F & $n,f_z$ & \sqrtf{\epsilon_\bk}{2(c_0+c_1)n}  & 0 & 0\\\cline{2-5}
         & $f_x,f_y$ & 0  & 0 &  \frac12 \\
\hline
P & $n$ &  \frac{ k\xi_n}{2} & 0 & 0  \\\cline{2-5}
         & $f_z$  & 0 & 0 & 0 \\\cline{2-5}
         & $f_x,f_y$ &  0 &  \frac{1}{2\sqrt{1+2c_1n/q}} & \frac{1}{2\sqrt{1+2c_1n/q}} \\
\hline
AF & $n$ & \left(1 - \frac12 \frac{f_z^2}{n^2} \frac{c_1}{c_0}  \right)\frac{k \xi_n }{2} &  \frac{f_z^2}{n^2}\sqrt{1-\frac{f_z^2}{n^2}}\frac{c_1^{3/2}}{c_0^{3/2}} \frac{k \xi_n }{2}  & 0 \\\cline{2-5}
         & $f_z$ & \frac{f_z^2}{n^2}\sqrtf{c_1}{c_0}  \frac{k \xi_s }{2}  & \sqrt{1-\frac{f_z^2}{n^2}}\left(1 - \frac{3 f_z^2}{2n^2} \frac{c_1}{c_0}  \right)\frac{k \xi_s }{2} &  0 \\\cline{2-5}
         & $\genfrac{}{}{0pt}{}{ f_x \to + }{f_y \to - }$ &  0 & 0 & \frac{1}{2}\left(\frac{1\pm\alpha_z}{1\pm\alpha_z-q/c_1 n} \right) \frac{E_{\mathrm{g},2}^{\mathrm{AF}}}{c_1 n} \\
\hline
BA & $n$  & \sqrtf{\efree}{2(c_0+c_1)n} &  0  & \frac{\qt^2}{(1-\qt^2)^{3/2}}\fracb{\efree}{c_1n}^2 \\\cline{2-5}
$(p=0)$         & $f_x$ & \frac{1}{1-\qt^2}\sqrtf{\efree}{2(c_0+c_1)n} & 0 &  \frac{\qt^2}{\sqrt{1-\qt^2}} \\\cline{2-5}
         & $f_y$ & 0 & \frac12(1+\qt)\sqrtf{q}{\efree} & 0 \\\cline{2-5}
         & $f_z$ & 0 &  \frac12(1-\qt) \sqrtf{\efree}{q} & 0 \\
\hline\hline
       \end{tabular}
   \caption{Small $k$ limits of fluctuation amplitudes. For $n$ and $f_z$ in the AF phase, the results are the first terms in an expansion for $c_1\ll c_0$. Where the entry is zero, the fluctuation amplitude is zero for all $k$.}
   \label{tab:flucamplimits}
   }
\end{table*}

A feature of spinor condensates is that additional continuous symmetries can be broken, leading to new Nambu-Goldstone modes, as is predicted to occur for the AF and BA phases. For the AF phase we found that the asymmetry in the nematic order of the condensate was revealed through the $f_x$ and $f_y$ fluctuations. In the BA phase we observed a divergence in the $f_y$ fluctuations associated with the spontaneous development of a transverse (axial-symmetry-breaking) magnetization. Our results show that this divergence arises from the Nambu-Goldstone magnon mode. 
Interestingly, such a divergence in fluctuations was not observed in our results for the AF phase, which also has a Nambu-Goldstone magnon branch. The reason is that for the AF phase the broken symmetry manifests only in the nematic order of the condensate, not in the spin order. Indeed, an immediate extension of our theory is to assess fluctuations of the nematic density,
\begin{equation}
    \hat{q}_{\alpha\beta}(\mathbf{x})=\hat{\boldsymbol{\psi}}^\dagger(\mathbf{x})\mathrm{Q}_{\alpha\beta}\hat{\boldsymbol{\psi}}(\mathbf{x}),
\end{equation}
as a  generalisation of Eq.~(\ref{eq:nematic_cond}). We find that for the AF state the fluctuations in ${q}_{xy}$ diverge for $k\to0$ due to both Nambu-Goldstone modes, with the magnon branch dominating. Because some of the techniques used to image the spin density are also sensitive to the nematic density (e.g.~see \cite{Stamper-Kurn2013R}),  the measurement of such fluctuations may also be possible in experiments.

Our analysis here has been for a uniform system, and several factors will become important in applying these results to the experimental regime. First, external trapping potentials cause the total density of the condensate to vary spatially and a full treatment of the trapped system would require a large-scale numerical solution of the Gross-Pitaevskii equation for the condensate and of the Bogoliubov-de Gennes equations for the quasiparticles. However, our analysis can be applied to this situation using the local density approximation, i.e.~we consider the gas to be homogeneous at each point in space using the local value of the condensate density. A discussion of the local density approximation in relation to the density response of a scalar condensate is presented in Ref.~\cite{Zambelli2000}. Second, in our analysis of the AF and BA phases we have assumed that the axisymmetry is broken uniformly over the entire system. For the case where the system forms domains of local broken axisymmetry our analysis will only apply to each domain (also see discussion in \cite{Murata2007a}).

\section*{Acknowledgments}   We thank Y.~Kawaguchi for providing feedback on the manuscript. We  acknowledge support by the Marsden Fund of New Zealand (contract UOO1220). 
\appendix
\section{Limits of quasiparticle amplitudes}

The large $k$ limits of the structure factors can be found directly from Eq.~\eqref{eq:Skinf}. For the small $k$ limits, we diagonalise a $6\times6$ matrix to give the quasiparticle amplitudes for each phase. For the F and P phases, the quasiparticle amplitudes are given in Sec.~5.2 of \cite{Kawaguchi2012R}, and for the BA phase they are given in \cite{Uchino2010a}. 
For the AF phase, the quasiparticle amplitudes of the gapped mode are given in \cite{Kawaguchi2012R}. The two ungapped modes have $[\bu_{\bk\nu}]_0=[\bv_{\bk\nu}]_0=0$ (where $[\mathbf{q}]_m$ is the $m$ component of the vector $\mathbf{q}$), so since $[\bm{\xi}^{\mathrm{AF}}]_0=0$, $\delta\fh_{x,\bk\nu}=\delta\fh_{y,\bk\nu}=0$ for these two modes. For $\nh$ and $\fh_z$, we need the differences (for $0<c_1\ll c_0$)

\begin{align}
    [\bu_{\bk0}-\bv_{\bk0}]_\pm &= \sqrt{1\pm \frac{f_z}{n}}\left[ 1\pm \frac{f_z}{n}\left(1\mp \frac{5f_z}{4n}\right)\frac{c_1}{c_0} \right]\frac{\sqrt{k \xi_n}}2, \\
    [\bu_{\bk1}\!-\!\bv_{\bk1}]_\pm &= \frac{\sqrt{1\mp \frac{f_z}{n}}}{(1-\frac{f_z^2}{n^2})^{1/4}}\!\!\left[\pm1 \!-\! \frac{f_z}{n}\!\left(\!1 \pm \frac{3f_z}{4n} \right) \frac{c_1}{c_0}\!  \right]\frac{\sqrt{ k\xi_s}}2,
\end{align}
which reduce to the result in \cite{Kawaguchi2012R} for $f_z=0$ and small $k$. 
We use Eq.~\eqref{eq:dQKtilde_def} to obtain the fluctuation amplitudes from the quasiparticle amplitudes, with results shown in Table~\ref{tab:flucamplimits}.
To get the small $k$ limits of the structure factors from the quasiparticle amplitudes, we use Eq.~\eqref{Eq:SQ}.
\bibliographystyle{apsrev4-1}

\end{document}